*Review*

# The Big World of Nanothermodynamics

**Ralph V. Chamberlin**

Department of Physics, Arizona State University, Tempe, AZ 85287-1504, USA
E-Mail: ralph.chamberlin@asu.edu; Tel.: +1-480-965-3922

**Abstract:** Nanothermodynamics extends standard thermodynamics to facilitate finite-size effects on the scale of nanometers. A key ingredient is Hill's subdivision potential that accommodates the non-extensive energy of independent small systems, similar to how Gibbs' chemical potential accommodates distinct particles. Nanothermodynamics is essential for characterizing the thermal equilibrium distribution of independently relaxing regions inside bulk samples, as is found for the primary response of most materials using various experimental techniques. The subdivision potential ensures strict adherence to the laws of thermodynamics: total energy is conserved by including an instantaneous contribution from the entropy of local configurations, and total entropy remains maximized by coupling to a thermal bath. A unique feature of nanothermodynamics is the completely-open nanocanonical ensemble. Another feature is that particles within each region become statistically indistinguishable, which avoids non-extensive entropy, and mimics quantum-mechanical behavior. Applied to mean-field theory, nanothermodynamics gives a heterogeneous distribution of regions that yields stretched-exponential relaxation and super-Arrhenius activation. Applied to Monte Carlo simulations, there is a nonlinear correction to Boltzmann's factor that improves agreement between the Ising model and measured non-classical critical scaling in magnetic materials. Nanothermodynamics also provides a fundamental mechanism for the $1/f$ noise found in many materials.

## 1. Introduction

Standard thermodynamics is strictly valid only for the homogeneous behavior of large systems, whereas thermal fluctuations often involve small systems. In 1962 Hill introduced the theory of small-system thermodynamics [1,2], establishing the fundamental laws that govern finite-size effects in thermodynamics. Although originally developed to describe individual molecules and isolated nanoparticles, this theory is also crucial for treating the heterogeneous distribution of independently relaxing regions that is now known to dominate the primary response of most materials [3-8]. The term "nanothermodynamics" was first published in the context of using small-system thermodynamics to treat nanometer-sized fluctuations inside bulk materials [9,10]; which is the focus of this brief review.

A key ingredient in nanothermodynamics is the subdivision potential ($E$) that must be included in the $1^{st}$ law of thermodynamics if total energy is to be conserved [11]. Hill's $E$ can be understood by comparison to Gibbs' chemical potential, $\mu$. $\mu$ is the change in energy to take a single particle from a bath of particles into the system, whereas $E$ is the change in energy to take a cluster of $n$ interacting particles from a bath of clusters into the system, and in general $n$ interacting particles do not have the same energy as $n$ isolated particles due to surface terms, length-scale effects, thermal fluctuations, etc. Thus, $E$ is needed to systematically treat the nonlinear and non-extensive contributions to energy from a system that contains a heterogeneous distribution of independent regions. Here I describe how the laws of nanothermodynamics guide the development of models and theories that treat independent thermal fluctuations inside bulk samples, and yield improved agreement with the measured response of many materials [9,12-17]. Indeed, these concepts provide a common explanation for several empirical features in the slow response of complex systems including: non-exponential relaxation, non-Arrhenius activation, non-classical critical scaling, and non-homogeneous response. As an introductory example I describe a fundamental mechanism for the non-Nyquist electronic noise found in many materials [18].

Nature exhibits several types of noise due to thermal fluctuations [19]. In 1827, Brown first reported sporadic motion of pollen grains in water. In 1905, as the second breakthrough in his *Annus Mirabilis*, Einstein explained this "Brownian motion" by assuming random impulses from much smaller particles, which was to provide the first definitive evidence for atoms and molecules. As a function of frequency ($f$) Brownian motion exhibits noise with a power spectral density that varies as $S(f) \propto 1/f^2$, In 1926 Johnson first measured electronic noise that showed a flat spectral density, $S(f) = $ const. Nyquist explained this "white" noise by assuming classical thermal fluctuations in the motion of the electrons. Also in 1926 Johnson measured electronic noise with apparent $1/f$ frequency dependence. Although there is still no widely accepted explanation for this "$1/f$ noise," empirically it is the most common low-frequency behavior. Indeed $1/f$ noise is found in metals, semimetals, semiconductors, dielectrics, ferroelectrics, ionic conductors, spin glasses, supercooled liquids, and quantum devices [20-23], as well as in music, speech, neural response, and human perception [24-27]. Although no single model is likely to explain all these phenomena, I use the laws of nanothermodynamics as a guide to obtain a unified picture for $1/f$ noise in many materials. Specifically, the general principle is based on the assumption that particles inside local regions of a bulk sample: conserve total energy by including non-extensive terms in $E$ ($1^{st}$ law), maintain maximum entropy during equilibrium fluctuations ($2^{nd}$ law), and/or exhibit statistical indistinguishability of identical particles consistent with quantum mechanics, as described in the next section.

**Figure 1.** From Ref. [18]. **(a)** Scaled magnetization versus time at two temperatures, from simulations of the Ising model. Note that simulations from lattices with different sizes, *N*, are offset for clarity. The dynamics changes abruptly at *t* = 0 when a nonlinear correction (Eq. (2)) is included with Boltzmann's factor (Eq. (1)). **(b)** Histograms of the simulations from *N* = 12 (symbols) and from the noise measured in three types of samples (solid lines). Note the trimodal behavior from the simulations, spin glass, and nanopore system.

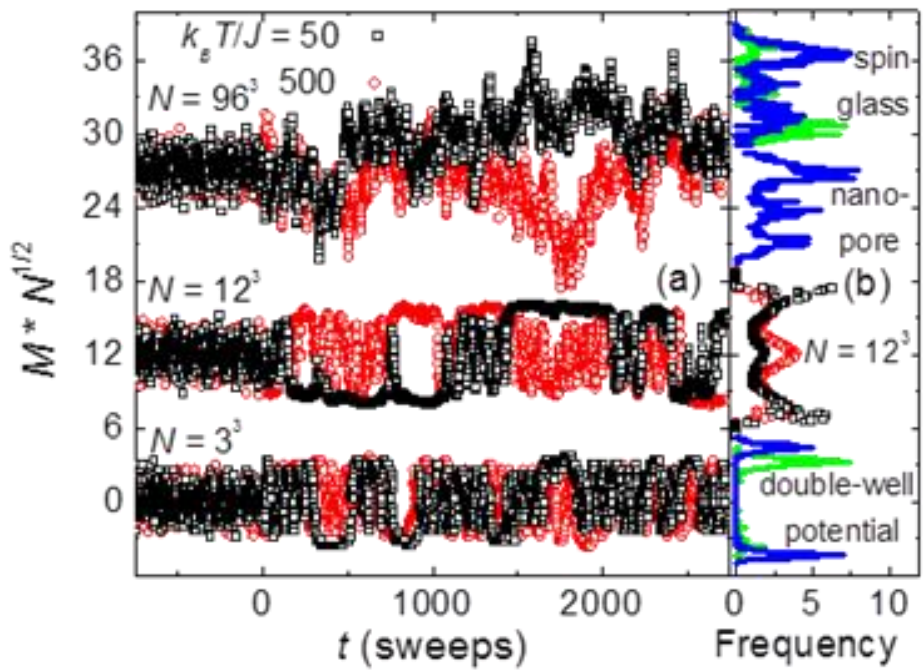

## 2. Nanothermodynamics in Computer Simulations

All simulations presented here utilize the Ising model for binary degrees of freedom ("spins") on a simple-cubic lattice. The lattice contains a total of *N* spins, with exchange interaction (*J*) between nearest-neighbor spins, and periodic boundary conditions between the outside surfaces of the lattice. Often the lattice is subdivided into smaller regions containing *n* < *N* spins to study the thermal properties of small systems inside a bulk sample. Figure 1 (a) shows the net magnetization as a function of time from simulations of this model. Note the abrupt change in dynamics at time *t* = 0: for *t* < 0 the spin-flip probability is governed by Boltzmann's factor alone using the Metropolis algorithm Eq. (1), whereas for *t* ≥ 0 there is also a nonlinear correction from nanothermodynamics Eq. (2).

$$e^{-\Delta U/k_B T} > rand[0,1) \quad \text{Boltzmann Factor} \tag{1}$$

$$e^{g(S_m - S_0)/k_B} > rand[0,1) \quad \text{Nonlinear Correction} \tag{2}$$

In Eq. (1), $\Delta U$ is the change in interaction energy, $k_B$ is Boltzmann's constant, and *T* is temperature. In Eq. (2), *g* controls the strength of the nonlinear correction that comes from the alignment entropy using the binomial coefficient for binary degrees of freedom: $S_m = k_BT \ln \dfrac{n!}{[\frac{1}{2}(n+m)]![\frac{1}{2}(n-m)]!}$, with $S_0 = k_BT \ln \dfrac{n!}{[(\frac{1}{2}n)!]^2}$ its maximum value. When $g = 0$ ($t < 0$ in Fig. 1(a)) systems of all size show standard Gaussian fluctuations characteristic of white noise. When $g=1$ ($t \geq 0$) the uppermost set of data (from a large lattice with small regions) shows large wandering on all time scales, indicative of $1/f$ noise; while the lower two sets of data (from smaller lattices that contain a single region) exhibit sharp jumps and steps characteristic of non-Gaussian fluctuations. Indeed, Fig. 1 (b) shows that histograms of the simulations exhibit trimodal behavior (symbols), similar to the trimodal behavior found from measurements of $1/f$ noise in a spin glass and ionic conduction through a nanopore, shown in the upper part of Fig. 1 (b) (solid lines). In contrast, the bottom pair of lines, from fluctuations in two different double-well potentials, shows simpler bimodal behavior.

**Figure 2.** From Ref. [18]. Left side shows frequency dependence of noise from simulations at $k_BT/J$=50 and 500, similar to those in Fig. 1. $S(f)$ is multiplied by $N$ to scale different lattice sizes (given in the legend) and log($f$) is multiplied by 10 to match the dB scale. The spectra exhibiting white noise (bottom) come from using Eq. (1) alone. Spectra exhibiting $1/f$–like behavior (diagonal) come from the same model using both Eq. (1) and Eq. (2). Over a broad range of frequencies these simulations can be characterized by $S(f) \propto 1/f^{\alpha(T)}$, with $\alpha(T) \approx 1.0$ for $k_BT/J$=500 (solid line) and $\alpha(T) \approx 1.15$ for $k_BT/J$=50 (dotted line). Diamond-shaped symbols, which show $1/f$ noise at low frequencies and white noise at higher frequencies, come from a heterogeneous system. Right side of figure shows $\alpha(T)$ from measurements on various metallic thin-films (solid symbols) and simulations (open hexagons connected by solid lines). Note that the temperatures are normalized by $T_1$, where $\alpha(T)$ extrapolates to 1.

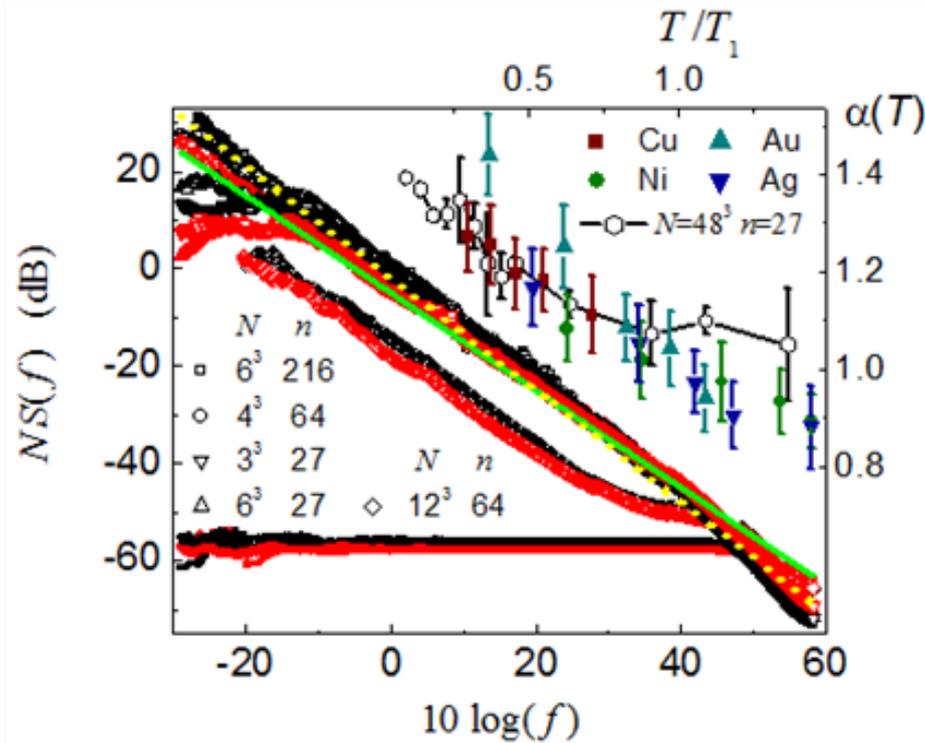

**Figure 3.** From Ref. [18]. (**a**) Sketch of possible states in a region containing two binary spins. (a) For distinguishable spins there is one way to have both up ($\Omega_{+2}$=1) or down ($\Omega_{-2}$=1), but two ways to have zero net alignment ($\Omega_0$=2). (**b**) During thermal fluctuations the Boltzmann entropy of the spins goes up and down (dashed line). To maintain maximum entropy the entropy of a bath must go down and up (dotted line), so that the combined entropy of the system plus bath is constant (solid line). (**c**) When the bath has high entropy each low-entropy state in the region persists twice as long as expected from the Boltzmann factor alone. (**d**) Alternatively, zero alignment may come from a single state that contains a superposition of spins, consistent with de-localized particles that are indistinguishable in the region.

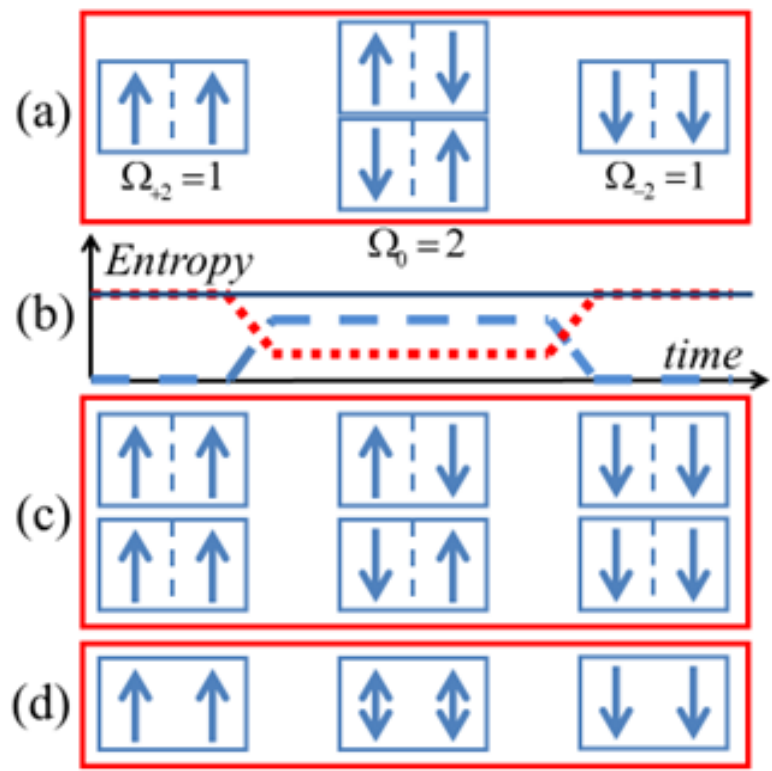

The left side of Fig. 2 shows the Fourier transform from simulations similar to those in Fig. 1 (a). Simulations using Eq. (1) alone yield white noise that does not depend on $f$ (lower symbols), whereas adding Eq. (2) yields $1/f$-like noise (symbols along the diagonal). In fact these simulations can be characterized by $\mathrm{S}(f) \propto 1/f^{\,\alpha(T)}$, with a temperature-dependent spectral-density exponent $\alpha(T)$ that models the measured behavior from several metals at lower temperatures, as shown in the upper part of Fig. 2. Similar large-amplitude low-frequency noise is found in most substances. Thus, Eq. (2) provides a formula for obtaining measured $1/f$-like noise, including deviations from pure $1/f$-behavior. Moreover, the nonlinear correction is based on fundamental physical laws from nanothermodynamics.

Figure 3 depicts some of the mechanisms that justify the nonlinear correction. Consider two binary degrees of freedom, e.g. two uniaxial spins that can be up or down. Figure 3 (a) shows that there is one way to have both spins up, one way to have both spins down, and two ways to have no net alignment. The alignment entropy is obtained from Boltzmann's definition, $S/k_{\mathrm{B}} = \ln(\Omega)$, using the multiplicity of each alignment, $\Omega$. The dashed (blue) line in Fig. 3 (b) shows how, during normal thermal fluctuations, this entropy may fluctuate up-and-down between its minimum $S/k_B = 0$ and maximum $S/k_B = \ln(2)$ values. Although seeming to violate the $2^{nd}$ law of thermodynamics, various explanations have been proposed. First: the entropy of small system may decrease temporarily [28], but clear violations of the $2^{nd}$ law requires a completely-closed system, which cannot be measured. Second: entropy could be defined using Gibbs ensembles that average over all available states, but this inhibits using entropy for time-dependent and out-of-equilibrium behavior. A third possibility is that when the entropy of a local region decreases, the entropy of its bath increases, so that the total entropy remains maximized. Indeed, Fig. 3 (b) shows how the entropy of the bath (dotted line) may balance the entropy of the system (dashed line), so that the combined entropy of the system plus bath is constant (solid line).

This mechanism for the nonlinear correction is a type of entropic force, similar to Boltzmann's factor [17]. Boltzmann's factor favors low-energy states because increasing the energy of the system lowers the entropy of the bath. Similarly, the nonlinear correction favors low-entropy states because increasing the alignment entropy of the system lowers the entropy of the bath. In fact, for the two-spin system the nonlinear correction precisely doubles the average lifetime of each fully aligned state, as shown schematically in Fig. 3 (c). Thus, each aligned state becomes as likely as both unaligned states. Information theory provides additional insight into this mechanism. If there is no transfer of information between the system and its environment, then the alignment of the system cannot be known and its multiplicity always includes all four states. Again the entropy is constant, but at a higher value $S/k_B=\ln(4)$.

A second mechanism for the nonlinear correction comes from the statistics of indistinguishable particles, as shown schematically in Fig. 3 (d). To match the probability of each alignment in Fig. 3 (c), instead of doubling the likelihood of the aligned states, the unaligned state could be a single superposition of up- and down-spins, as expected for identical particles that are subject to the exchange interaction. Indeed, the three net alignments (up, down, and unaligned) form the triplet states of a two-spin system. Further justification for this interpretation comes from the energy and its fluctuations shown in Fig. 4 (below), where a similar nonlinear correction minimizes the energy of small regions and makes their entropy extensive and additive, consistent with the statistics of indistinguishable particles. Thus, this mechanism for extensive entropy in small regions is similar to the semi-classical ideal gas that resolves Gibbs' paradox for extensive entropy in large volumes. In other words, the nonlinear correction may be a simplistic way to simulate quantum-like behavior in classical models.

A third way to understand the nonlinear correction is from conservation of energy including Hill's subdivision potential. Consider a system of $n$ non-interacting Ising-like spins with magnetic moment $\mu_B$ in field $B$. Each spin can be aligned or anti-aligned with $B$, giving energy $-\mu_B B$ or $+\mu_B B$, respectively. The single-particle partition function is $Z_1 = e^{\mu_B B/k_B T} + e^{-\mu_B B/k_B T}$. Because the spins are non-interacting, the partition function for the entire system is $Z = (Z_1)^n$, yielding the free energy $A = -nk_B T \ln(Z_1)$. For simplicity let $B \rightarrow 0$, so that $A = -nk_B T \ln(2)$. Again using the binomial coefficient for the exact entropy of the system, the internal energy becomes $U_m = A + TS_m = -nk_B T \ln(2) + k_B T \ln \dfrac{n!}{[\frac{1}{2}(n+m)]![\frac{1}{2}(n-m)]!}$.

If the system is at its average alignment $m = 0$, Stirling's approximation for the factorials yields $S_0 \approx n$ $\ln(2)$ and $U_0 \approx 0$. However, if total energy is to be conserved during fluctuations, there is a non-extensive contribution to internal energy $U_m = U_0 - E_m$, where $E_m = k_B T (S_0 - S_m)$ is Hill's subdivision potential. In other words, although $U_0 \approx 0$ at $m = 0$, during thermal fluctuations the change in alignment entropy causes a change in energy, independent of any interaction between particles. Fluctuations in $E_m$ occur because free energy is a thermal average in the canonical ensemble, while energy and entropy may be defined in each microcanonical state. Note that when $m \neq 0$, $E_m > 0$ lowers the total energy, which favors subdividing a bulk sample into an ensemble of regions that fluctuate independently to increase the fluctuations, as is found in the primary response of most materials. Inserting this $E_m$ from entropy into a Boltzmann-like factor yields the nonlinear correction, Eq. (2). Interaction energies that appear explicitly in Eq. (1) neglect this source of energy, so that Eq. (2) is essential if total energy is to be strictly conserved in finite-sized systems.

The mismatch between nanothermodynamics and standard simulations using Eq. (1) alone is due to finite-size effects from assuming homogeneous systems with uniform correlations. Specifically, when energy fluctuations are averaged over a volume that excludes interacting particles outside the volume, correlations across the interface are neglected. For large homogenous systems the nonlinear correction may give a negligible contribution to conservation of energy: $n \rightarrow \infty$ yields $m \rightarrow 0$, and $E_0 = 0$. However, several experimental techniques have established that the primary response of most materials comes from a heterogeneous distribution of independently-relaxing nanometer-sized regions. Indeed, dynamical heterogeneity in the correlations between interface particles has been found in the slow response of liquids, glasses, polymers, and crystals [3-8]. In fact, extensive studies of time-dependent specific heat at low temperatures find that the only substance to show no evidence for localized excitations is chemically and isotopically pure NaF crystals [29]. Furthermore, the technique of nonresonant spectral hole burning (NSHB) establishes that excess energy is localized in selected degrees of freedom inside a sample for minutes, or even hours [30,31], without influencing the energy in neighboring regions. Thus, complexity in the primary response of most substances comes from thermodynamic heterogeneity due to an ensemble of independently relaxing regions, consistent with nanothermodynamics, but different from the behavior assumed for standard thermodynamics.

Energy fluctuations in most Monte Carlo simulations exhibit a size dependence [32] that differs from the expectation that entropy is extensive and additive. Figure 4 shows the average potential energy density ($<U/J>/n$) and its fluctuations ($<(\Delta U/kT)^2>/n$) as a function of the number of particles ($n$) in local regions of a large lattice. Again the simulations utilize the standard Ising model for binary spins on a simple-cubic lattice with ferromagnetic interaction between nearest-neighbors. The solid (black) lines, from simulations using Boltzmann's factor alone (Eq. (1)), show constant energy density, whereas the fluctuations in energy density increase with increasing region size. The size dependence of these energy fluctuations yields a size-dependent specific heat, and non-extensive entropy that is incompatible with the laws of nanothermodynamics for an ensemble of independently relaxing regions. The symbols in Fig. 4 show the energy density and its fluctuations for the same Ising model with various strengths ($g$) in an approximation of Eq. (2) that yields a quadratic correction to Boltzmann's factor, with a bypass.

$$\exp\left[-g\,\tfrac{1}{2}\,\frac{m^2}{n}(1 - \delta_{\Delta U,0})\right] \qquad \text{Quadratic Correction} \qquad (3)$$

**Figure 4.** From Ref. [15]. (**a**) Average energy per particle and (**b**) its fluctuations from MC simulations of the Ising model  Solid lines come from using the standard Metropolis algorithm; symbols come from simulations with varying strength of a nonlinear correction, with *g*=1 the expected correction from a Taylor series expansion of entropy. The dashed lines show that *g*=1 yields energy reductions proportional to 1/*n* in (a), and constant energy fluctuations in (b). (**c**) Average energy per particle, and (**d**) its fluctuations, as a function of the strength of the nonlinear correction. Each type of symbol comes from a different region size, as given in the legend.

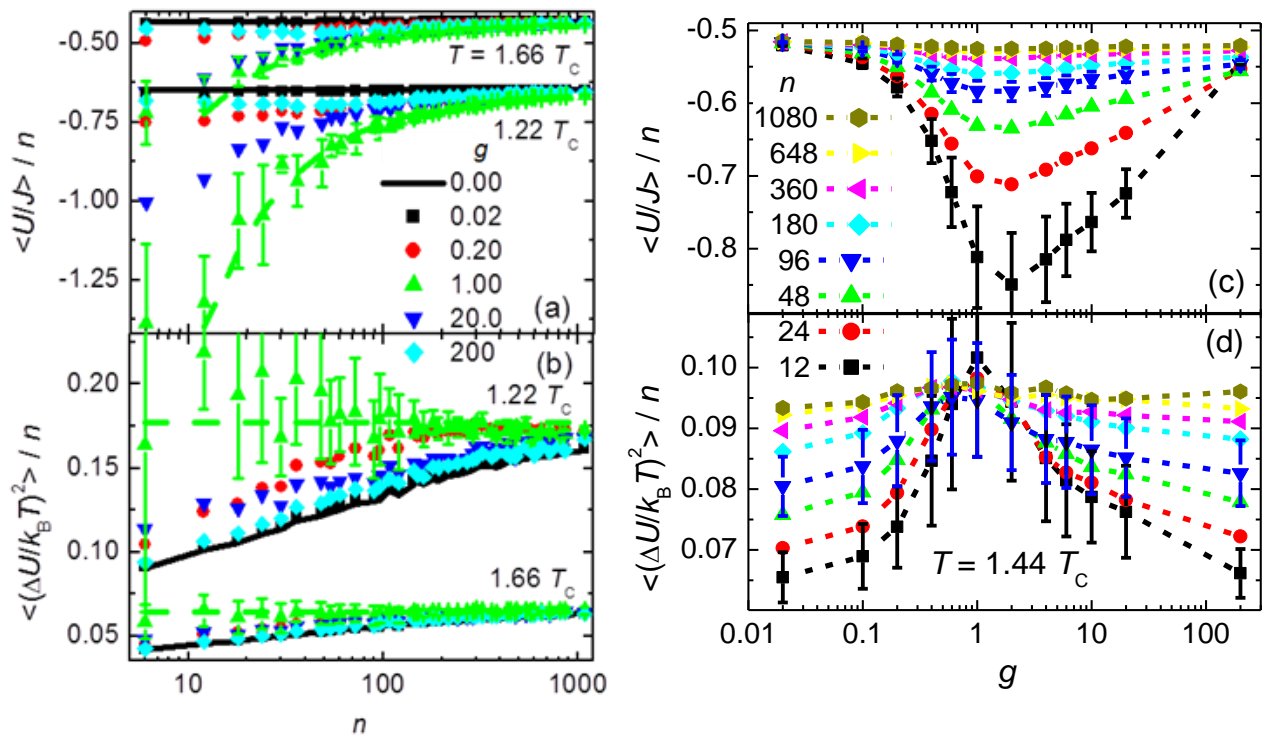

The bypass comes from the Kronecker delta that gives $(1-\delta_{\Delta U,0})=0$ when $\Delta U = 0$. A practical reason for this bypass is to accelerate slow relaxation and avoid frozen response. A physical reason is that if $\Delta U = 0$ spin flips can occur in a microcanonical ensemble, without coupling to the heat bath. Sufficient disorder in a local region may remove the bypass if neighboring states with $\Delta U = 0$ are not available, consistent with the fact that 1/*f* noise generally increases with increasing disorder. Furthermore, most materials exhibit a combination of 1/*f* noise at low frequencies and white noise at high frequencies, indicating dynamics without, and with, the bypass, respectively. Indeed, the diamond-shaped symbols in Fig. 2 come from a heterogeneous mixture of Eq. (2) and $e^{(S_m-S_0)(1-\delta_{\Delta U,0})/k_B} > rand[0,1)$. In any case, Figs. 4 (a)-(d) show that increasing *g* in Eq. (3) reduces the energy density and increases the energy fluctuations per particle, until *g*=1 where the energy of small regions is minimized and the specific heat is independent of region size. Thus, *g*=1 yields behavior that is most consistent with thermodynamic equilibrium: energy that is minimized and fluctuations in energy that yield extensive entropy. Empirical evidence for extensive entropy, even in small regions inside bulk samples, comes from the technique of NSHB that shows constant specific heat for independently relaxing regions [33], even for very small regions having *n* ~ 10 molecules [34].

**Figure 5.** From Ref. [16]. (**a**) Critical scaling plot of magnetic susceptibility versus reduced temperature, $\tau = (T\text{-}T_C)/T_C$. Symbols are from measurements of gadolinium. Lines are from simulations of the Ising model with ($g$=1, solid) and without ($g$=0, dashed) the nonlinear correction. (**b**) Residual plot of the data and $g$=0 simulations, from $g$=1 at the origin. (**c**) Effective scaling exponent (magnitude of slope from Fig. 5 (**a**)) for the Gd data and Ising-model simulations.

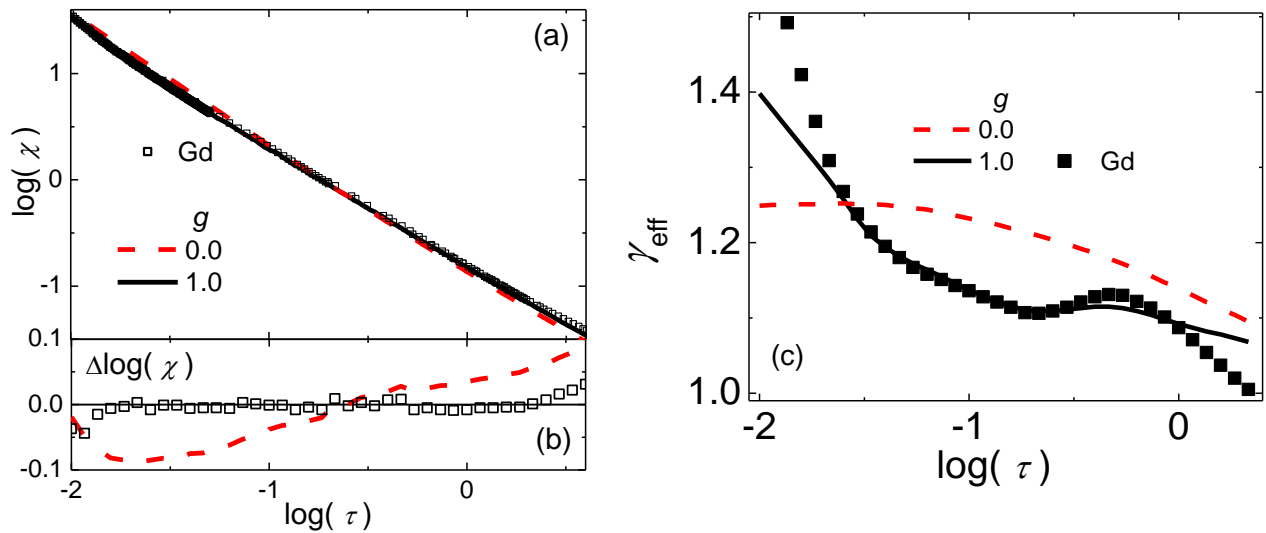

## 3. Comparison of Computer Simulations with Measurements

The nonlinear correction also improves agreement between computer simulations and the measured response of many materials. The symbols in Fig. 5 (a) show the magnetic susceptibility from single-crystal gadolinium as a function of reduced temperature. At high temperatures the data have a slope of −1.0, consistent with the Curie-Weiss law. As $T \rightarrow T_C$, the red (dashed) line from simulations of the Ising model using Boltzmann's factor alone shows a slope of −1.24, consistent with the expectation from renormalization group theory for the homogeneous Ising model in the canonical ensemble. However, close inspection of the data reveals that the slope is not constant as $T \rightarrow T_C$. The solid (black) curve in Fig. 5 (a), which shows improved agreement with the data at all temperatures, comes from simulations of the same Ising model with the nonlinear correction (Eq. (2)) expected for heterogeneous systems obeying nanothermodynamics. Figure 5 (b) shows the difference between the data (symbols) and simulations of the Ising model without (dashed line), and with (solid line at the origin), the nonlinear correction. Indeed, using $g$=1 with the optimal region size ($n$=27) reduces the standard deviation by an order of magnitude. Figure 5 (c) shows the effective scaling exponent $\gamma_{eff}$ as a function of reduced temperature, from the magnitude of the slope when plotted as in Fig. 5 (a). The data (symbols) and simulations with nonlinear correction (solid line) show conspicuous features that clearly differ from the monotonic behavior [35] of simulations using Boltzmann's factor alone (dashed line). In fact, when measured to within 0.01 % of $T_C$, ultra-high-purity gadolinium crystals show a sharp maximum with $\gamma_{eff} > 1.5$, and $\gamma_{eff} \rightarrow 1$ as $T \rightarrow T_C$ [36], consistent with mean-field theory at the transition.

**Figure 6.** From Ref. [17]. Solid line shows the excess specific heat as a function of normalized temperature from measurements of LaMnO$_3$ around its Jahn-Teller distortion. Symbols and dashed line come from simulations of the Ising model with and without the nonlinear correction.

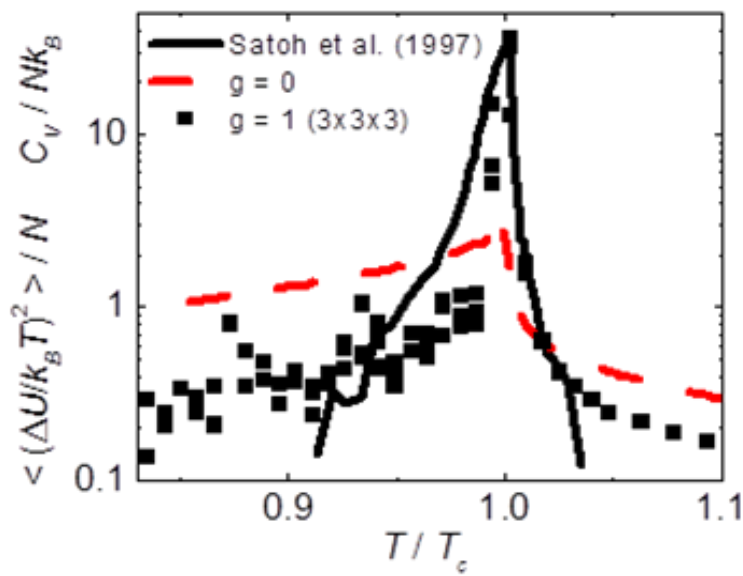

The solid line in Fig. 6 shows the excess specific heat measured in the colossal magneto-resistance material LaMnO$_3$. The peak identifies the phase transition from a Jahn-Teller distortion that occurs at about 730 K. The dashed line comes from simulations of the Ising model using Boltzmann's factor alone ($g$=0), while the solid symbols come from the same model with the nonlinear correction ($g$=1) and the optimal region size of $n$=27. Again the nonlinear correction gives significantly better agreement with the measured behavior.

Direct evidence for heterogeneous correlations in crystals of LaMnO$_3$ comes from neutron scattering [37,38]. The upper two lines in Fig. 7 show the pair distribution function (PDF) measured above, and below, the Jahn-Teller transition temperature. The difference between these PDFs (lower curve) shows strong correlations in the positions of neighboring atoms out to a distance of ~1.0 nm, then an abrupt loss in correlation, with a more-gradual loss in correlation beyond this abrupt jump. Strong correlations that persist for three lattice sites are consistent with $n = 3^3$ used for the simulations in Fig. 6. The clear non-monotonic loss in correlation is incompatible with the classical Ornstein-Zernicke picture [39], where the pair-correlation function is predicted to diminish smoothly and homogeneously around every atom. Thus, the classical picture of homogeneous correlations is invalid on length scales of nanometers, where quantum mechanics may influence the correlations.

**Figure 7.** From Refs. [17,38]. Pair distribution functions (PDFs) from neutron scattering in LaMnO$_3$. Upper lines show two PDFs; one from above the Jahn-Teller distortion temperature and one from below, with an identical background removed from both. Lower line shows the difference between these two PDFs. Note the enhanced correlation at short distances, then abrupt loss of correlation at radius $r \approx 1.0$ nm, corresponding to distance between three lattice sites.

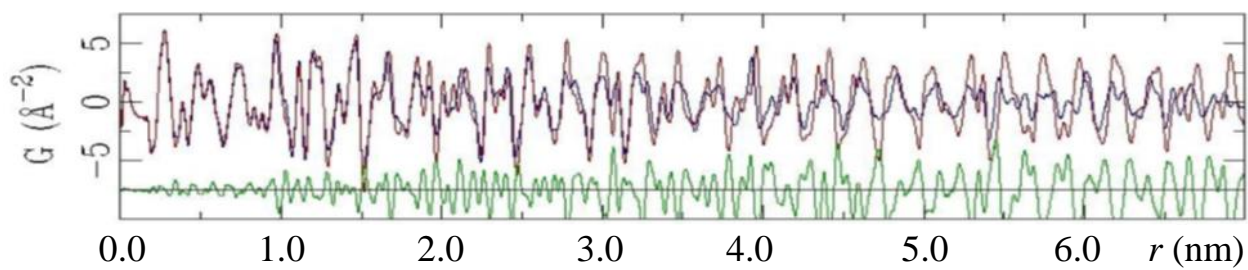

**Figure 8.** From Refs. [17,38]. Pair correlation function versus distance, from simulations of the Ising model without (*g*=0) and with (*g*=1) the nonlinear correction [17]. Inset shows inverse correlation length from neutron scattering (circles), and *g*=1 simulations (squares, with error bars) found from linear fits to the data over the 3 unit cells within a region.

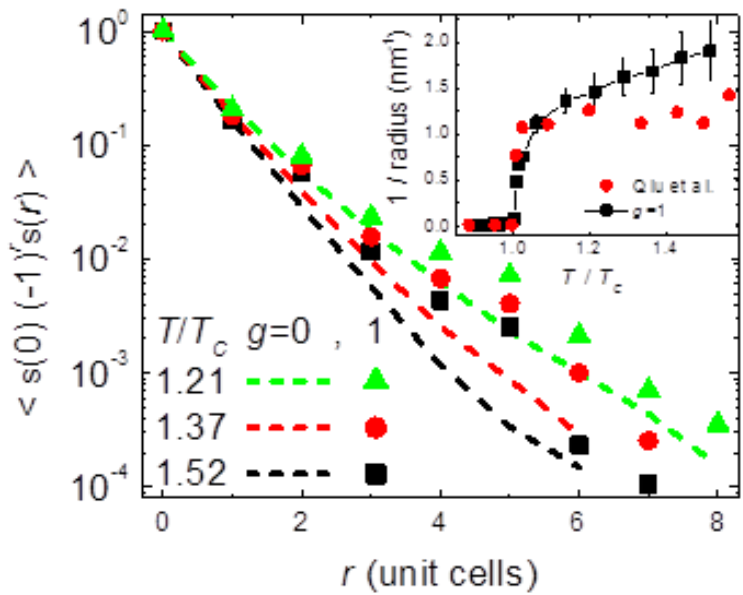

Figure 8 shows the pair-correlation function from simulations of the Ising model without (*g*=0, dashed lines) and with (*g*=1, symbols) the nonlinear correction. When *g*=0 there is a smooth loss in correlation, characteristic of the classical picture. When *g*=1 there is stronger correlation over the three contiguous lattice sites within each region, then an abrupt loss in correlation to the neighboring region, consistent with the measured pair distribution function shown in the lower part of Fig. 7. The inset of Fig. 8 shows that linear regression on three adjacent sites in a region, from simulations with *g*=1, yields a correlation length that is similar to the radius of distinct regions determined by neutron scattering.

Additional direct evidence for heterogeneous correlations comes from multi-dimensional NMR [40,41]. Measurements and analysis yield an average size for the independently relaxing regions of 10, 76, and 390 molecules (or monomer units) in glycerol, ortho-terphenyl, and poly(vinyl acetate), respectively. The distribution of relaxation times gives response rates that can vary by several orders of magnitude across nanometer length scales. Nonresonant spectral hole burning measurements indicate that this heterogeneity corresponds to effective local temperatures that also vary between these regions, indicating *thermo*dynamic heterogeneity. Thus, accurate models and theories of nanometer-sized independently-relaxing regions must obey the laws of nanothermodynamics.

**Figure 9.** From Ref. [13]. Various ensembles for fluctuations inside a bulk sample. The microcanonical ensemble applies to fully-closed fluctuations that conserve number of particles (*n*), volume (*v*), and energy (*ε*). The canonical ensemble (*n,v,T*) applies to fluctuations at constant volume when heat flows in and out from the rest of the sample. The grand canonical ensemble (*μ,v,T*) applies to fluctuation at constant volume when particles also flow in and out from the rest of the sample. The "nanocanonical" ensemble (*μ,P,T*) applies to fully-open fluctuations, where the volume of the fluctuation is allowed to change as particles flow in and out.

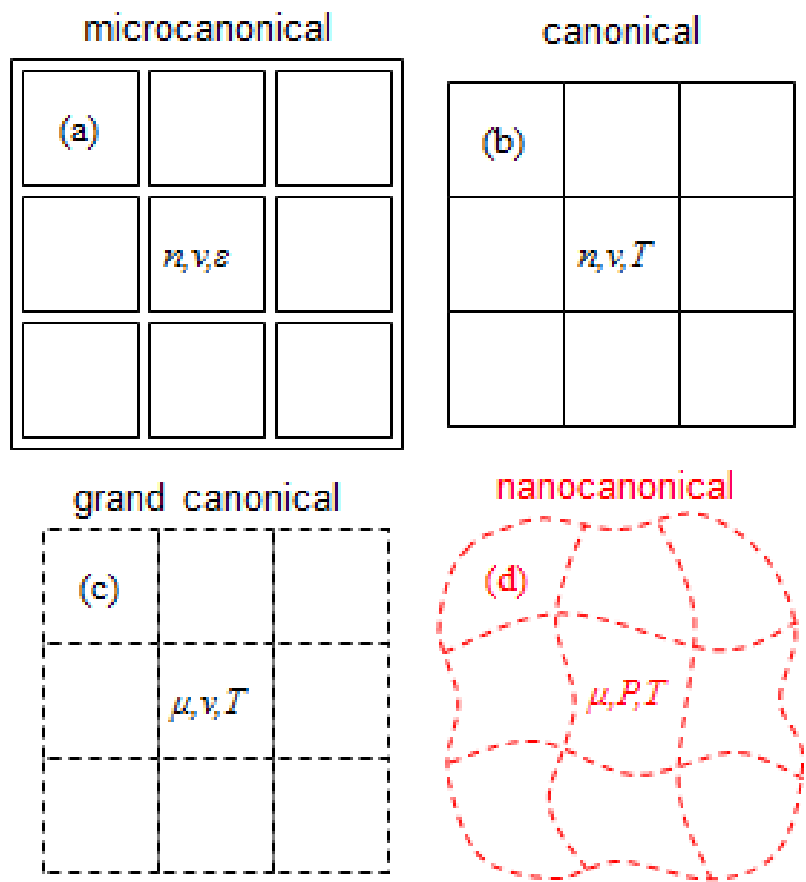

## 4. Nanocanonical Ensemble, Equilibrium Landau Theory, and Mean-Field Cluster Model

Figure 9 depicts various thermodynamic ensembles for independent fluctuations inside bulk samples. Figure 9 (a) shows a fast fluctuation that does not have time to couple to its environment, so that the internal energy (*ε*), volume (*v*), and number of particles (*n*) of the fluctuation are conserved, yielding the microcanonical ensemble. Figure 9 (b) shows a slower fluctuation that allows heat to pass in and out, so that *ε* fluctuates and *T* replaces *ε* as an environmental variable, yielding the canonical ensemble. Figure 9 (c) shows a fluctuation where particles may join and leave the fluctuation, so that *n* fluctuates and *μ* replaces *n* as an environmental variable, yielding the grand canonical ensemble. Figure 9 (d) shows a fully-open fluctuation where density is conserved as particles join and leave the fluctuation, so that *v* fluctuates and pressure (*P*) replaces *v* as an environmental variable, yielding Hill's "generalized" ensemble. This *μ, P, T* ensemble might also be called "nanocanonical" in accordance with the other ensembles. The nanocanonical ensemble is ill-defined in standard thermodynamics.

Indeed, the Gibbs-Duhem equation yields zero free energy for the nanocanonical ensemble, requiring at least one extensive environmental variable to control the size of the systems. For small systems, however, the subdivision potential accommodates non-extensive contributions to energy that allow the small systems to control their own size without external constraints. In fact, NSHB measurements show that some energy remains localized in slow degrees of freedom for minutes, or even hours. Thus, the most relevant ensembles for independently fluctuating regions inside bulk samples are the microcanonical ensemble for fast fluctuations that do not have time to couple to their environment, and the nanocanonical ensemble for slow fluctuations that must couple fully to their environment without artificial constraints. Furthermore, because different ensembles yield different dynamics, finite-sized fluctuations should be treated using the correct ensemble.

The nanocanonical ensemble forms the basis of the mean-field cluster model and equilibrium Landau theory for phase transitions. In Landau theory, the symmetry of the system is used to obtain an energy per particle that depends quadratically on the order parameter (e.g. net alignment) of the region $\varepsilon(m) = -\frac{1}{2}\varepsilon_2 m^2$, where $\varepsilon_2$ is a constant parameter. When combined with the entropy, approximated to fourth order, the free energy per particle becomes $f(m) = \varepsilon(m) - TS_m/m \approx f_0 + \frac{1}{2}(k_B T - \varepsilon_2) m^2 + k_B T m^4/12$, where $f_0 = -k_B T \ln(2)$. For simplicity the integral over all possible energies is extended to $m = \pm\infty$, yielding the canonical partition function $Z_n = \int_{-\infty}^{\infty} e^{-n f(m)/k_B T} dm$. In the usual thermodynamic limit, the system size is assumed to be very large $n \to \infty$, so that the system remains at a minimum in Helmholtz free energy, found from $\partial f(m)/\partial m = 0$. Whereas in the nanocanonical ensemble, region size is also allowed to fluctuate, minimizing the free energy for fluctuations without artificial size constraints. The partition function is $\Upsilon = \sum_{n=n_0}^{\infty} Z_n e^{n\mu/k_B T}$, where $n_0$ is a minimum size for thermal behavior. The mean-field cluster model improves upon the nanocanonical Landau theory by using exact expressions for the alignment entropy of each region $S_m$, and replacing the integral in $Z_n$ by a sum over allowed alignments. In both cases, $\mu/k_B T$ is adjusted to the constant value that gives best agreement with temperature-dependent data; similar to, but simplified from, an ideal gas where $\mu/k_B T$ depends logarithmically on temperature.

Figure 10 shows paramagnetic susceptibility ($\chi$) as a function of reduced temperature above the ferromagnetic transition, $T > T_C$. The symbols are from measurements on four standard ferromagnetic materials, as given in the legend. The classical mean-field theory (Curie-Weiss law) for homogeneous systems predicts a constant critical-scaling exponent of $\gamma = 1$ in $\chi \propto 1/(T - T_C)^\gamma$, whereas the data show a steeper slope, especially around $(T - T_C)/T_C \approx 0.01$. Nearer $T_C$, microscopic models that assume homogeneous correlations via the canonical ensemble predict a constant slope with scaling exponents of $\gamma > 1$, whereas close inspection of the data show temperature-dependent curvature. The solid curves are from the mean-field cluster model, with $\mu/k_B T$ as the parameter governing the temperature-dependent shape of the susceptibility. Even in the raw data it is possible to see that the mean-field cluster model for a heterogeneous distribution of independently relaxing regions gives better agreement than homogeneous models. Figure 10 (b) shows the effective scaling exponent $\gamma_{eff}$ as a function of reduced temperature for two sets of data from Fig. 10 (a). Again these data are clearly incompatible with the Ising model in the canonical ensemble (dashed line), and again more-recent measurements show that ultra-pure Gd crystals never have a constant non-classical scaling exponent other than $\gamma_{eff} = 1$

**Figure 10.** From Ref. [9]. (**a**) Critical-scaling plot of paramagnetic susceptibility as a function of reduced temperature. The symbols show measurements of four standard ferromagnetic materials, as given in the legend, using data in the literature from several laboratories. The solid curves are from a mean-field cluster model in the nanocanonical ensemble, with chemical potential as the key adjustable parameter that controls the shape of the curves. (**b**) Effective scaling exponent, from the numerical derivative of two sets of data (symbols) and model curves (solid lines) from (a). Also shown is the behavior of the standard Ising model from computer simulations using the Metropolis algorithm (Eq. 1) that assumes homogeneous correlations and the canonical ensemble.

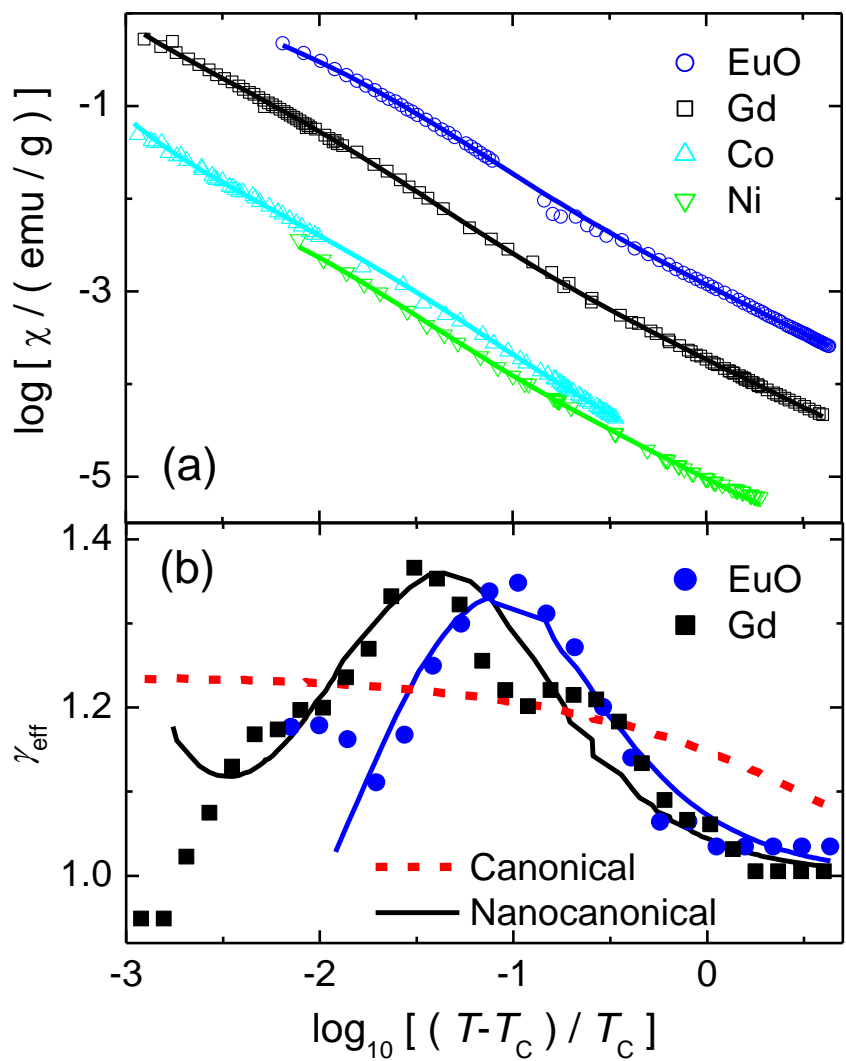

[42], consistent with mean-field theory in the nanocanonical ensemble (solid lines). Indeed, the nanocanonical ensemble significantly improves agreement between mean-field models and measured response, similar to how the nonlinear correction from nanothermodynamics improves agreement with computer simulations of the microscopic version of the same model, as shown in Figs. 5 and 6.

**Figure 11.** Temperature dependence of inverse peak-loss frequency (left axis) and inverse magnetic susceptibility (right axis), plotted in a way that yields constant slopes at high temperatures for the Vogel-Fulcher and Curie-Weiss laws, respectively. Curvature indicates deviations from these laws. Symbols show measurements of two glass-forming liquids (left) and two ferromagnetic materials (right). Solid lines are from the mean-field cluster model.

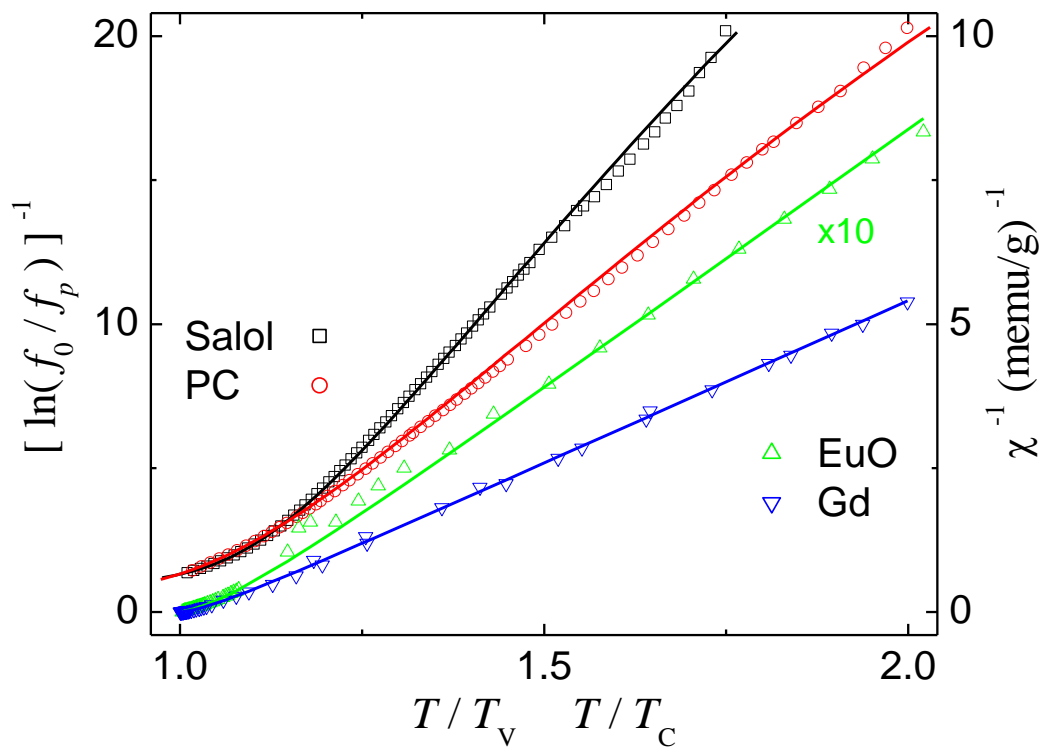

The lower two sets of symbols in Fig. 11 show inverse magnetic susceptibility versus temperature from two ferromagnetic materials. The data are some of the same as shown in Fig. 10, but on a linear scale to emphasize the high-temperature regime. At $T > 1.2T_C$ the data follow a straight line, consistent with the Curie-Weiss law $1/\chi \propto (T−\theta)$, where $\theta$ is the Weiss temperature. Again as $T{\rightarrow}T_C$ there is curvature, consistent with the solid curves from the mean-field cluster model for non-classical critical scaling. The upper two sets of symbols in Fig. 11 show the temperature dependence of the peak-loss frequency $f_p$, from measurements of dielectric susceptibility of two glass-forming liquids. The data are plotted in a way that yields a straight line if the systems obey the VFT law for super-Arrehenius activation $1/\ln(f_0/f_p) \propto (T−T_V)$, where $T_V$ is the Vogel temperature. Similar behavior is characteristic of the WLF formula for scaling of the thermal and dynamic properties of polymers [14]. Note the mathematical similarity between the VFT and Curie-Weiss laws, and the qualitatively similar curvature as the critical temperatures are approached. Indeed, VFT-like behavior and measured deviations can also be characterized by the mean-field cluster model, as shown by the solid curves. In glass-forming liquids the underlying phase transition is more subtle, due to larger fluctuations in smaller regions that broaden the transition, and dynamical slowing as the transition is approached. Figure 12 is an Angell plot of $1/f_p$ from several glass-forming liquids (symbols) that show VFT-like behavior. Again the solid lines show that the VFT law and measured deviations can be characterized by the mean-field cluster model for a phase transition in a nanocanonical distribution of independently relaxing regions.

**Figure 12.** From Ref. [12]. Angell plot of inverse peak frequency versus inverse temperature. The Arrhenius law yields straight lines on this plot. The symbols are from six different supercooled liquids, as given in the legend. The solid curves are from the mean-field cluster model, with constant $\mu/k_BT$ as the key parameter that governs the shape of the curves.

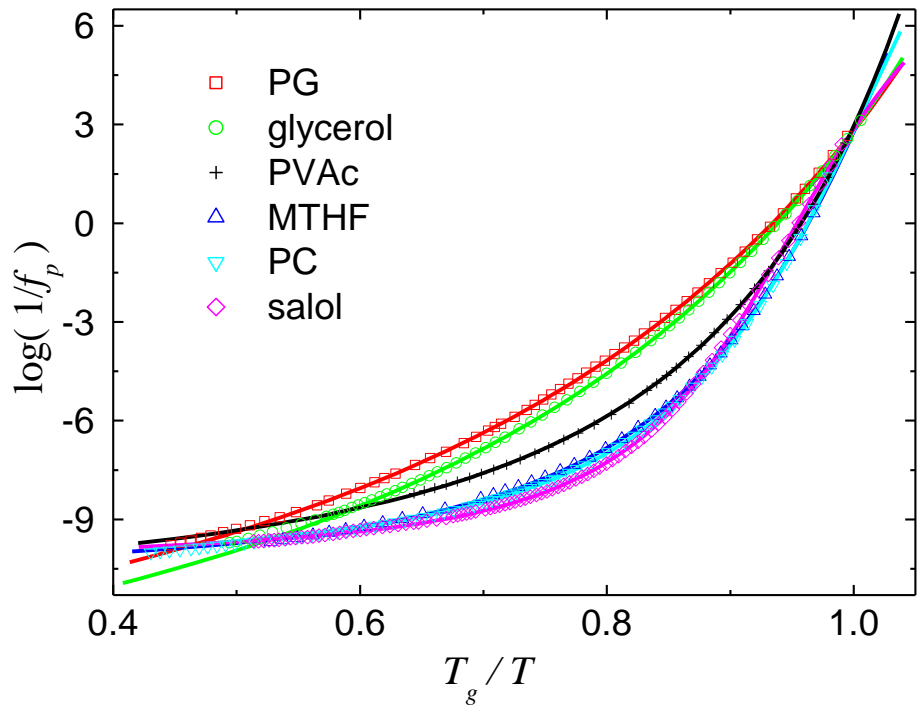

Nanothermodynamics is needed to obtain the heterogeneous distribution of independently relaxing regions inside bulk samples, found from minimizing the free energy in the nanocanonical ensemble. Of course the size of individual regions is dynamic, fluctuating with time to increase the total entropy. However, in thermal equilibrium the average size and distribution are well defined. Connecting the nanocanonical distribution of region sizes to the equilibrium spectrum of response requires a size-dependent relaxation rate. Good agreement with data is obtained using relaxation rates that vary exponentially with inverse size, $w_n = w_\infty \exp(C_V/2nk_B)$. Here $w_\infty$ is the asymptotic relaxation rate for large regions and $C_V$ is the heat capacity for the relaxing degrees of freedom in an average-sized region. The inverse-size dependence in $w_n$ may be related to the $1/n$ energy reduction due to the nonlinear correction, as shown by the dashed lines in Fig. 4 (a). The $C_V/n$ dependence in $w_n$ signifies the importance of thermal fluctuations: large regions have large heat capacity so that they fluctuate less (per particle) than small regions, reducing the likelihood that large regions will fluctuate between localized states, slowing their relaxation. Indeed, the factors of $N$ used to match the amplitudes in Figs. 1 and 2 confirm that the mean-squared fluctuations vary inversely proportional to size. The specific form of $w_n$ can be found using a model where fluctuations cause an overlap between discrete energy levels [43]. In any case, the inverse-size dependence in the relaxation rate yields good agreement with measured spectra of response.

**Figure 13.** From Ref. [12]. Log-log plot of dielectric loss as a function of frequency. The symbols are from measurements on glycerol at four temperatures. Solid lines are from the mean-field cluster model, with the heat capacity of an average-sized region $C_V$ as the adjustable parameter that governs the width and shape of the spectra. The dashed line shows a constant slope of magnitude β=0.57, characteristic of the KWW law for stretched-exponential relaxation.

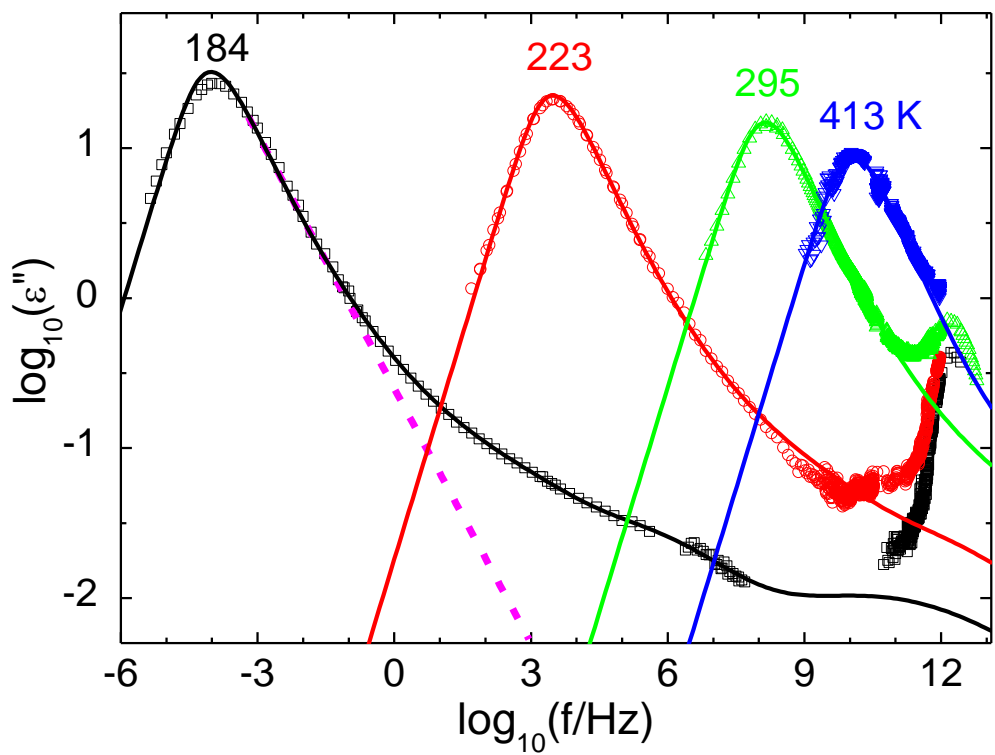

The symbols in Fig. 13 show the frequency-dependent dielectric loss of glycerol at four different temperatures. The solid curves come from $w_n$, with the distribution of sizes given by thermal equilibrium of the nanocanonical ensemble. The solid curves show good agreement with several observed features. The asymmetric slopes come from the inverse-size dependence in $w_n$, yielding +1 on the low-frequency side of the peak and −β on the high-frequency side, similar to the KWW law for stretched-exponential relaxation shown by the dashed line. The high-frequency wing comes from small regions, when $C_V/n > k_B$. The step-like features that start at the highest frequencies come from integer values of $n = 1, 2$, and perhaps 3. For $n > 3$ the response from discrete regions merges into a smooth curve. Note that the widths of the dielectric-loss peaks are nearly constant over a wide range of temperatures. This approximate time-temperature superposition is incompatible with a static distribution of activation energies, indicating that the response involves thermal fluctuations that cancel the explicit temperature dependence in the Arrhenius law [14], consistent with $w_n$. Also note that $w_n$ predicts that the width and spectrum of response are governed primarily by the specific heat of an average-sized region. Indeed, Fig. 14 shows that the temperature dependences of $C_V$ deduced from dielectric-loss spectra agree with the values found directly from measurements of excess specific heat.

**Figure 14.** From Ref. [12]. Specific heat as a function of temperature. Solid symbols come from $C_V$ deduced from fits to dielectric loss spectra of glycerol (from Fig. 13), salol, and propylene carbonate. Open symbols (connected by solid lines to guide the eye) come from direct measurements of excess specific heat.

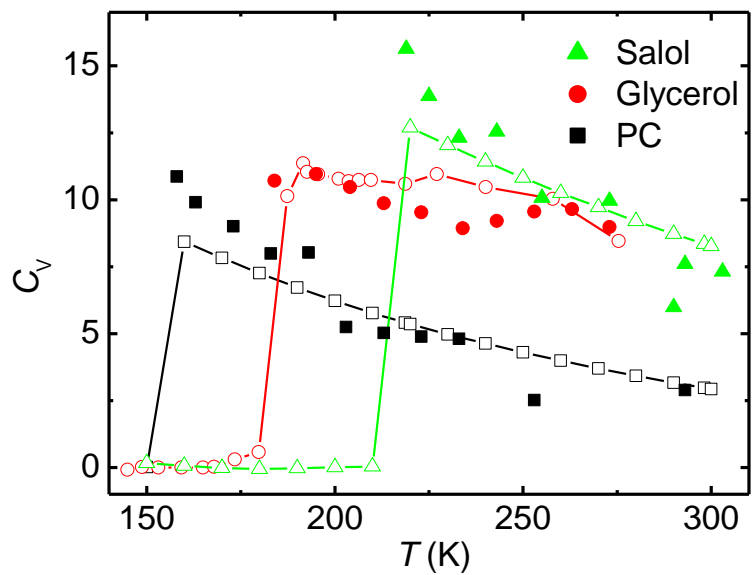

## 5. Discussion

Some remarkably universal empirical formulas have been used to characterize the measured response from many materials. The Kohlrausch-Williams-Watts law is used for stretched-exponential relaxation, as shown in Fig. 13. The Vogel-Fulcher law is used for super-Arrhenius activation, as shown in Figs. 11 and 12. Non-classical critical-scaling exponents are used for behavior near phase transitions, as shown in Figs. 5, 10, and 11. Non-Nyquist fluctuations, having a frequency dependence of $1/f^\alpha$ with $\alpha \sim 1$, are used for the electronic noise in most materials, as shown in Figs. 1 and 2. In many cases these expressions are convenient mathematical formulas for cataloging the measured response of complex systems, but data of sufficient quality over broad enough range invariably show deviations from these formulas. Many models have been proposed for each of the empirical formulas, so that the deviations may be a decisive way to distinguish between models. Nanothermodynamics provides a common foundation for all of these empirical formulas, including many of the measured deviations. Moreover, nanothermodynamics is necessary to describe the thermal equilibrium of any sample that contains independently fluctuating regions if the fluctuations themselves are to govern the distribution of region sizes.

Of course nanothermodynamics is not a universal theory of everything, there must be a mechanism that allows independently fluctuating regions inside the sample. For example, thermodynamic heterogeneity is not expected for the coherent ground state of superfluids and superconductors. Indeed, non-classical critical scaling from the homogeneous XY model provides extraordinary agreement with measured specific heat near the lambda transition in $^4$He [44]. Similarly, simulations using nanothermodynamics differ from Onsager's solution of the Ising model, because the canonical ensemble in an infinite sample with homogeneous correlations is incompatible with a heterogeneous distribution of independently relaxing regions. Another example is Tsallis entropy that has been used to characterize the properties of various systems [45]. In fact, some form of non-extensive entropy [46] is necessary to explain the behavior of Monte Carlo simulations that use Boltzmann's factor alone, as shown by $g=0$ in Fig. 4 where the energy fluctuations in small regions are non-extensive. Also, the $1/f$ characteristics found in most music, markets, and human perception may have a psychological, not physical basis; but perhaps human preferences are influenced by our environment.

Standard models based on homogeneous thermodynamics have been unable to explain several features in the dynamics of complex systems. The deviations may be quite subtle. Indeed, it is difficult to see curvature in the data on a log-log critical-scaling plot, as shown by Figs. 5 (a) and 10 (a). Nevertheless, other researchers have also recognized that most ferromagnetic materials deviate from standard critical-scaling behavior. In 1989 Collins wrote [47]: "The critical exponents of iron and nickel are very similar to each other, while those for cobalt are clearly different. There is no theoretical understanding of these results." Also in 1989 Hohenemser et al. wrote [48]: "At the same time our review makes clear that when one restricts the analysis to the best experiments, only a few materials correspond unambiguously to these models, while most do not." In fact, by plotting the residuals (as in Fig. 5 (b)) there is obvious improvement between measured behavior and the Ising model when treated using concepts from nanothermodynamics. Moreover, the monotonic behavior of standard simulations of the Ising model cannot match the sharp temperature-dependent features in the effective scaling exponent, as shown in Figs. 5 (c) and 10 (b). In any case, the nanocanonical ensemble must be used if independently-fluctuating regions inside bulk samples are to be included in the thermal equilibrium.

## 6. Conclusions

Nanothermodynamics is not a specific model, it is a general principle. No such principle can fully describe the detailed behavior of any system. However, like any thermodynamics it establishes the fundamental laws for what is physically possible. When data are found to deviate from a standard model, instead of searching for more complicated models it may be useful to first adapt the simpler model to obey the laws of nanothermodynamics, as shown by the improved agreement between the Ising model and measured behavior in Figs. 1, 2, 5, 6, and 8. From the discussion around Fig. 3 it is possible to emphasize the following laws of nanothermodynamics. The $1^{st}$ law requires that total energy is strictly conserved, including Hill's subdivision potential from the configurational entropy in regions as they fluctuate. Strictly obeying the $2^{nd}$ law requires that the entropy of an isolated system must never decrease, so that if a local region fluctuates into a low entropy state, the entropy of its thermal bath must increase to compensate, thereby maintaining maximum entropy. Thus, nanothermodynamics is essentially an extension of standard thermodynamics to finite-sized systems, with strict adherence to the standard laws. Moreover, these laws yield the statistics of indistinguishable particles within each region, as needed to avoid non-extensive entropy, resolve Gibbs' paradox, and agree with quantum-mechanical behavior for fundamental particles in nanometer-sized regions. Indeed, the cooperative dynamics within each region and uncorrelated dynamics between neighboring regions, combined with the fact that standard thermodynamics accurately describes macroscopic coherent states, suggests a connection to quantum mechanics [49,50]. Thus, each region may be associated with localized excitations that are incoherent with the excitations in neighboring regions, so that nanothermodynamics may provide a fundamental connection between quantum mechanics on the scale of nanometers and the bulk behavior of most materials.

## Acknowledgments

This review is dedicated to the memory of Terrell L. Hill (1917–2014). Among his many other contributions to science and society he: laid the foundation for nanothermodynamics, originated the name, and helped guide me through the initial stages of its application to fluctuations inside bulk materials.

I thank D. Bedeaux, N. Bernhoeft, B.F. Davis, A. Shevchuk, R. Richert and G.H. Wolf for several stimulating discussions. I also thank S.J.L. Billinge, P. Lunkenheimer, M. Oguni, R. Richert and S. Srinath for providing much of the original data shown here. This research was supported by the ARO, W911NF-11-1-0419.

## Conflicts of Interest

The author declares no conflict of interest.